\documentclass[aps,prc,reprint,superscriptaddress]{revtex4-1}
\usepackage{graphicx}

\begin{document}

\title{Completing the nuclear reaction puzzle of the nucleosynthesis of  $^{92}$Mo}

\author{G. M. Tveten}
\email[]{g.m.tveten@fys.uio.no}
\affiliation{Department of Physics, University of Oslo, NO-0316 Oslo, Norway}
\author{A. Spyrou}
\affiliation{National Superconducting Cyclotron Laboratory, Michigan State University, East Lansing, Michigan 48824, USA}
\affiliation{Department of Physics and Astronomy, Michigan State University, East Lansing, Michigan 48824, USA}
\affiliation{Joint Institute for Nuclear Astrophysics, Michigan State University, East Lansing, Michigan 48824, USA}

\author{R. Schwengner}
\affiliation{Helmholtz-Zentrum Dresden-Rossendorf, 01328 Dresden, Germany}

\author{F. Naqvi}
\affiliation{National Superconducting Cyclotron Laboratory, Michigan State University, East Lansing, Michigan 48824, USA}
\affiliation{Joint Institute for Nuclear Astrophysics, Michigan State University, East Lansing, Michigan 48824, USA}

\author{A. C. Larsen}
\affiliation{Department of Physics, University of Oslo, NO-0316 Oslo, Norway}

\author{T. K. Eriksen}
\affiliation{Department of Physics, University of Oslo, NO-0316 Oslo, Norway}
\affiliation{Department of Nuclear Physics, Research School of Physics and Engineering, The Australian National University, Canberra ACT 2601, Australia}

\author{F. L. Bello Garrote}
\affiliation{Department of Physics, University of Oslo, NO-0316 Oslo, Norway}

\author{L. A. Bernstein}
\affiliation{Lawrence Livermore National Laboratory, Livermore, California 94551, USA}

\author{D. L. Bleuel}
\affiliation{Lawrence Livermore National Laboratory, Livermore, California 94551, USA}

\author{L. Crespo Campo}
\affiliation{Department of Physics, University of Oslo, NO-0316 Oslo, Norway}

\author{M. Guttormsen}
\affiliation{Department of Physics, University of Oslo, NO-0316 Oslo, Norway}

\author{F. Giacoppo}
\affiliation{Department of Physics, University of Oslo, NO-0316 Oslo, Norway}
\affiliation{Helmholtz Institute Mainz, 55099 Mainz, Germany}
\affiliation{GSI Helmholtzzentrum f\"{u}r Schwerionenforschung, 64291 Darmstadt, Germany}

\author{A. G\"{o}rgen}
\affiliation{Department of Physics, University of Oslo, NO-0316 Oslo, Norway}

\author{T. W. Hagen}
\affiliation{Department of Physics, University of Oslo, NO-0316 Oslo, Norway}

\author{K. Hadynska-Klek}
\affiliation{Department of Physics, University of Oslo, NO-0316 Oslo, Norway}
\affiliation{INFN, Laboratori Nazionali di Legnaro Padova, Italy}

\author{M. Klintefjord}
\affiliation{Department of Physics, University of Oslo, NO-0316 Oslo, Norway}

\author{B. S. Meyer}
\affiliation{Department of Physics and Astronomy, Clemson University, Clemson, SC 29634, USA}

\author{H. T. Nyhus}
\affiliation{Department of Physics, University of Oslo, NO-0316 Oslo, Norway}

\author{T. Renstr\o m}
\affiliation{Department of Physics, University of Oslo, NO-0316 Oslo, Norway}

\author{S. J. Rose}
\affiliation{Department of Physics, University of Oslo, NO-0316 Oslo, Norway}

\author{E. Sahin}
\affiliation{Department of Physics, University of Oslo, NO-0316 Oslo, Norway}

\author{S. Siem}
\affiliation{Department of Physics, University of Oslo, NO-0316 Oslo, Norway}

\author{T. G. Tornyi}
\affiliation{Department of Nuclear Physics, Research School of Physics and Engineering, The Australian National University, Canberra ACT 2601, Australia}

\date{\today}

\begin{abstract}
One of the greatest questions for modern physics to address is how elements heavier than iron are created in extreme, astrophysical environments. A particularly challenging part of that question is the creation of the so-called p-nuclei, which are believed to be mainly produced in some types of supernovae. The lack of needed nuclear data presents an obstacle in nailing down the precise site and astrophysical conditions.

In this work, we present for the first time measurements on the nuclear level density and average $\gamma$ strength function of $^{92}$Mo. State-of-the-art p-process calculations systematically underestimate the observed solar abundance of this isotope. Our data provide stringent constraints on the $^{91}$Nb$(p,\gamma)^{92}$Mo reaction rate, which is the last unmeasured reaction in the nucleosynthesis puzzle of $^{92}$Mo. Based on our results, we conclude that the $^{92}$Mo abundance anomaly is not due to the nuclear physics input to astrophysical model calculations.

\end{abstract}


\maketitle
\section{Introduction}
The observed distribution of heavy element abundances in our solar system provides a fingerprint of a complex interplay between nuclear properties and extreme, astrophysical environments. Our understanding and identification of the  stellar forges creating elements heavier than iron has improved significantly since the first attempts at understanding stellar nucleosynthesis in the 1950's. However, there are still mysteries regarding the astrophysical sites as well as the nuclear data needed to describe the heavy-element nucleosynthesis \cite{Arnould20031, Rauscher2013}. 

Perhaps one of the most intriguing remaining mysteries concerns the 35 stable isotopes that cannot be explained by the slow or rapid neutron-capture processes \cite{Arnould20031, Rauscher2013}. The so-called p-process was suggested as an explanation for the existence of these isotopes \cite{synthesis1957}.  As of today, $\gamma$-induced photodisintegration of preexisting seed nuclei  is understood to be the main production mechanism of the p-process \cite{Arnould20031,Rauscher2013} (also known as  the $\gamma$-process for this reason).

Favorable conditions for the p-process are found in the O-Ne layer of type II supernovae \cite{under2} and in type Ia supernovae \cite{under9}. Astrophysical model calculations are able to reproduce abundance patterns of most p-isotopes reasonably well, with some pivotal exceptions. In particular, p-isotopes of mass $92\leq A \leq98$ are underproduced in calculations compared to the actual abundance of these isotopes  \cite{under1,under2,under3,under4,under5,under6,under7,under8,under9,under10}. It has been suggested that the reason is related to the p-process seed nuclei as discussed in Ref. \cite{Arnould20031}. The underproduction could also be related to the details of the astrophysical site description. Experimental constraints on nuclear reaction rates are important to rule out the anomaly being related to the nuclear physics input.

\begin{figure}[tb]
\includegraphics[width=0.25\textwidth]{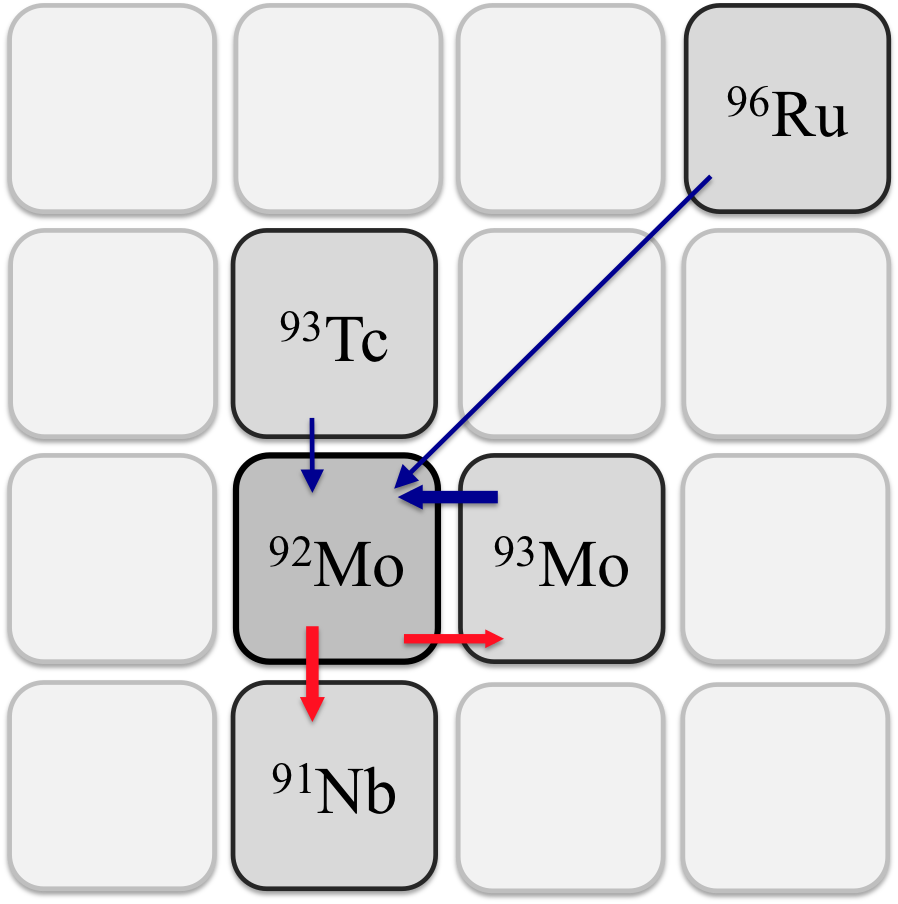}
\caption{The production and destruction mechanisms for $^{92}$Mo that are known to contribute $\gtrsim 1 \%$ to the final abundance. \label{fig:fig0}}
\end{figure}

In this work, we focus on one of the most severe cases: the underestimate of the abundance of $^{92}$Mo, which is typically underproduced by $1-2$ orders of magnitude \cite{Rauscher2013}. The production and destruction mechanisms are shown in Fig.\ref{fig:fig0} (figure adapted from Ref.\cite{figref}). Data constrain the reaction rates of $^{92}$Mo$(\alpha,\gamma)$ \cite{92Moag},  $^{92}$Mo$(p,\gamma)$ \cite{SauterPRC553127,Hasper2010,Gyurky2014112} and $^{92}$Mo$(n,\gamma)^{93}$Mo \cite{92mong, exfor}. The only reaction remaining as a possible source for the $^{92}$Mo puzzle is the dominant destruction reaction $^{92}$Mo$(\gamma,p)^{91}$Nb. It has been shown \cite{psensitivity} that the photodisintegration cross section of ${}^{92}$Mo($\gamma,$p)${}^{91}$Nb and the inverse reaction have a large impact on  the final abundances in p-process network calculations. Usually $(\gamma,p)$ cross sections are calculated from $(p,\gamma)$ cross sections by applying the reciprocity theorem \cite{reciprocity}, but  $^{91}$Nb is unstable making it challenging to use as target material.
We report the first experimental constraint on the cross-section, and consequently the astrophysical rate, for the $^{91}$Nb$(p,\gamma)^{92}$Mo reaction. We present new data for two of the most important nuclear input for capture cross-section calculations, namely the nuclear level density (NLD) and the  $\gamma$-ray strength function ( $\gamma$SF). The NLD represents the available number of quantum levels per section of excitation energy, $E_x$, while the $\gamma$SF is a measure of the  $\gamma$-absorption and decay properties for a given  $\gamma$-ray energy $E_{\gamma}$. We have applied the Oslo method \cite{firstgeneration,unfolding, Schiller2000498, omsystematics, oclsoft} to $^{92}$Mo$(p,p^{\prime}\gamma)^{92}$Mo data to extract the experimental NLD and $\gamma$SF of $^{92}$Mo for excitation energies up to the neutron separation energy. Further, we have used our data as input in Hauser-Feshbach \cite{hauser} calculations for extracting the first experimental constraint of the $^{91}$Nb$(p,\gamma)^{92}$Mo reaction rate. 

\section{Experimental procedure and data analysis}
The experiment was carried out at the Oslo Cyclotron Laboratory (OCL). A 16.5 MeV proton beam was directed at a self-supporting target of isotopically enriched ${}^{92}$Mo of $\approx 2$ mg/cm${}^{2}$ thickness, populating excited states in ${}^{92}$Mo through the $(p,p')$ reaction. The proton energies were measured with SiRi, a composite detector system consisting of eight trapezoidal-shaped silicon $\Delta E-E$ telescopes. The modules consist of a 1550 $\mu$m thick E detector with a 130 $\mu$m thick $\Delta$E detector in front \cite{Guttormsen2011168}. The $\Delta E$ detectors are segmented into 8 curved strips ($\Delta \theta = 2^{\circ}$) covering scattering angles between 126$^{\circ}$ and 140$^{\circ}$. Signals from SiRi open a time gate and $\gamma$-rays were measured in coincidence mode with the $5"\times$ 5"  NaI(Tl) scintillator $\gamma$-detector array CACTUS \cite{cactus}. Events were selected by gating on the $\Delta E-E$ curve corresponding to protons and reaction kinematics were used to calculate the excitation energy of $^{92}$Mo. Finally, the measured data were arranged in an ($E_{\gamma}, E_x$) coincidence matrix resulting in excitation-energy tagged $\gamma$-ray spectra for all $E_x$ bins. 

The $\gamma$-ray spectra were unfolded using the technique described in Ref. \cite{unfolding} with recently remeasured response functions \cite{unfolding,feprl}. The shape of the primary $\gamma$-ray spectra for each excitation-energy bin was determined from the iterative subtraction technique described in Ref. \cite{firstgeneration}, referred to as the first generation method. 
\begin{figure}[tb]
\includegraphics[ trim={0 0 0 1cm},width=0.5\textwidth]{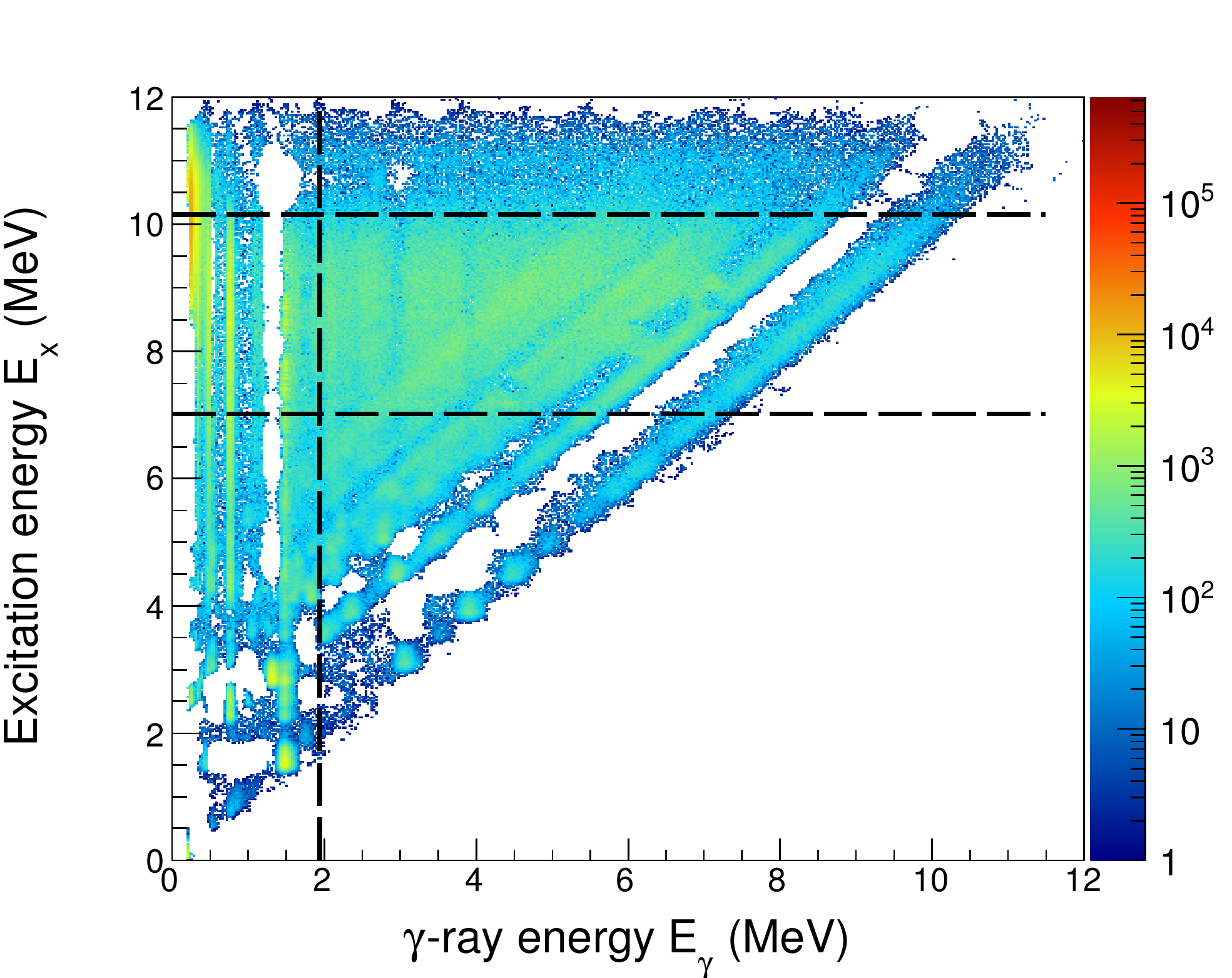}
\caption{The primary $E_x-\gamma$-ray matrix, $P(E_{\gamma},E_x)$, with the limits for the extraction of the of NLD and $\gamma$SF shown as dashed lines. \label{fig:fgplot}}
\end{figure}
Further, the functional shape of $\rho(E)$ and the transmission coefficient, $\mathfrak{T}(E)$, for  ${}^{92}$Mo were extracted simultaneously from the $E_x$-primary $\gamma$-ray energy matrix shown in Fig. \ref{fig:fgplot} for 7 MeV $\leq E_x \leq$ 10.2 MeV using the least square method described in Ref. \cite{Schiller2000498}. The lower limit on excitation energy was set to exclude non-statistical contributions from the  $P(E_{\gamma},E_x)$ matrix. The threshold for the (p,2p)-channel is 7.540 MeV and at 10.2 MeV the contribution of this channel becomes significant as was seen from the fluctuations in $\gamma$-multiplicity. Gamma ray energies $E_{\gamma} < 1.94$ MeV were also excluded, because the strong $2^{+} \rightarrow 0^{+}$ transition (higher-generation transition) present in the decay cascades was not removed properly in the first generation method. The statistical part of the normalized $P(E_{\gamma},E_x)$-matrix is assumed to be described by 
\begin{equation} \label{eq:eq0}
P(E_{\gamma},E_x) \propto \rho(E_x - E_{\gamma}) \mathfrak{T}(E_{\gamma}).
\end{equation}  
The resulting $\rho(E_x - E_{\gamma})$ and $\mathfrak{T}(E_{\gamma})$ reproduces the experimental primary spectra well, as shown in Fig. \ref{fig:itdoeswork} for selected excitation energies.
\begin{figure*}[tb]
\includegraphics[width=0.95\textwidth]{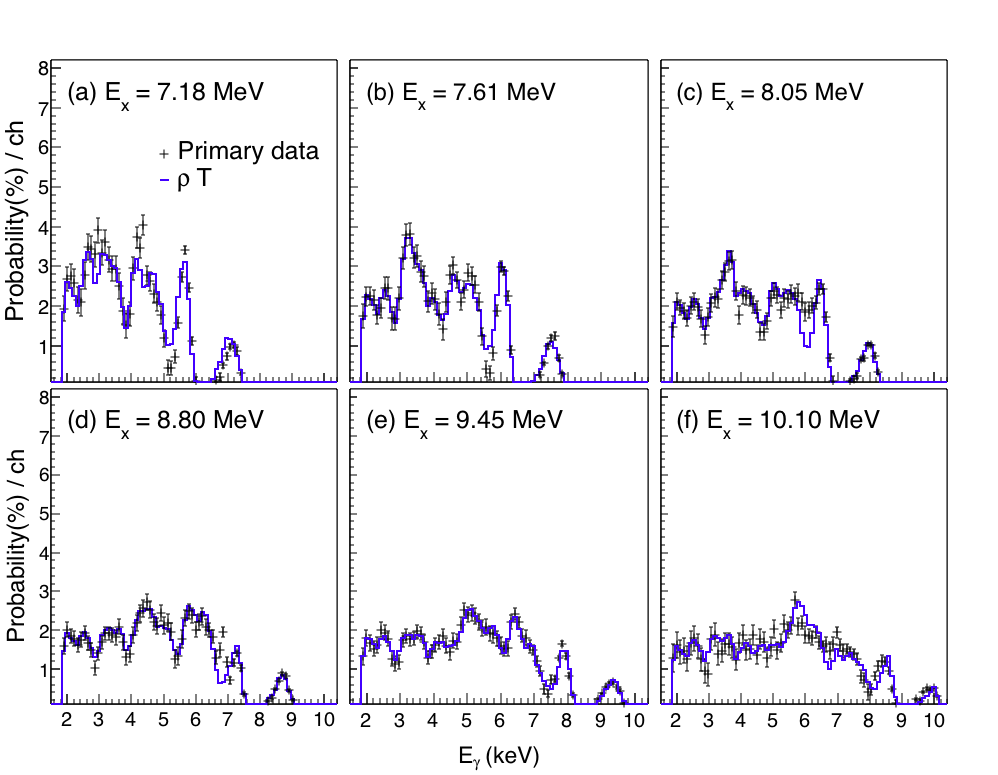}
\caption{(Coloronline) First-generation spectra from selected initial energies $E_i$(crosses) compared to the product of the level density, $\rho(E_i - E_{\gamma}$, and transmission coefficient vectors, $\mathfrak{T}(E_{\gamma} )$. The spectra are normalized to unity.\label{fig:itdoeswork}}
\end{figure*}

The absolute value and slope of $\rho(E)$ were determined from discrete levels \cite{endf} below an excitation energy of $E_x = 3$ MeV and from the level density at the neutron separation energy, $\rho(S_n)$. Since ${}^{91}$Mo, is an unstable isotope the normalization values at $S_n$ were estimated from systematics of level spacings from neighbouring isotopes \cite{PhysRevC.88.041305,PhysRevC.88.015805, Capote20093107}. 

The parity distribution of states is assumed to be symmetrical in the decaying energy region for the normalization of both the NLD and $\gamma$SF. According to the microscopic Hartree-Fock-Bogoliubov plus combinatorial calculations of Ref. \cite{ghk08} the parity distribution should be rather symmetric for $E_x \gtrsim 6$ MeV. Experimentally, no parity dependence was observed for the case of $^{90}$Zr that has a similar nuclear structure to $^{92}$Mo \cite{Kalmykov07}. However, even in the case of parity asymmetry being present, it was shown in Ref. \cite{omsystematics} that the contribution is modest. The large part of the uncertainty in this analysis is due to the uncertainty in normalization values at $S_n$.

\begin{figure}[tb]
\includegraphics[width=0.5\textwidth]{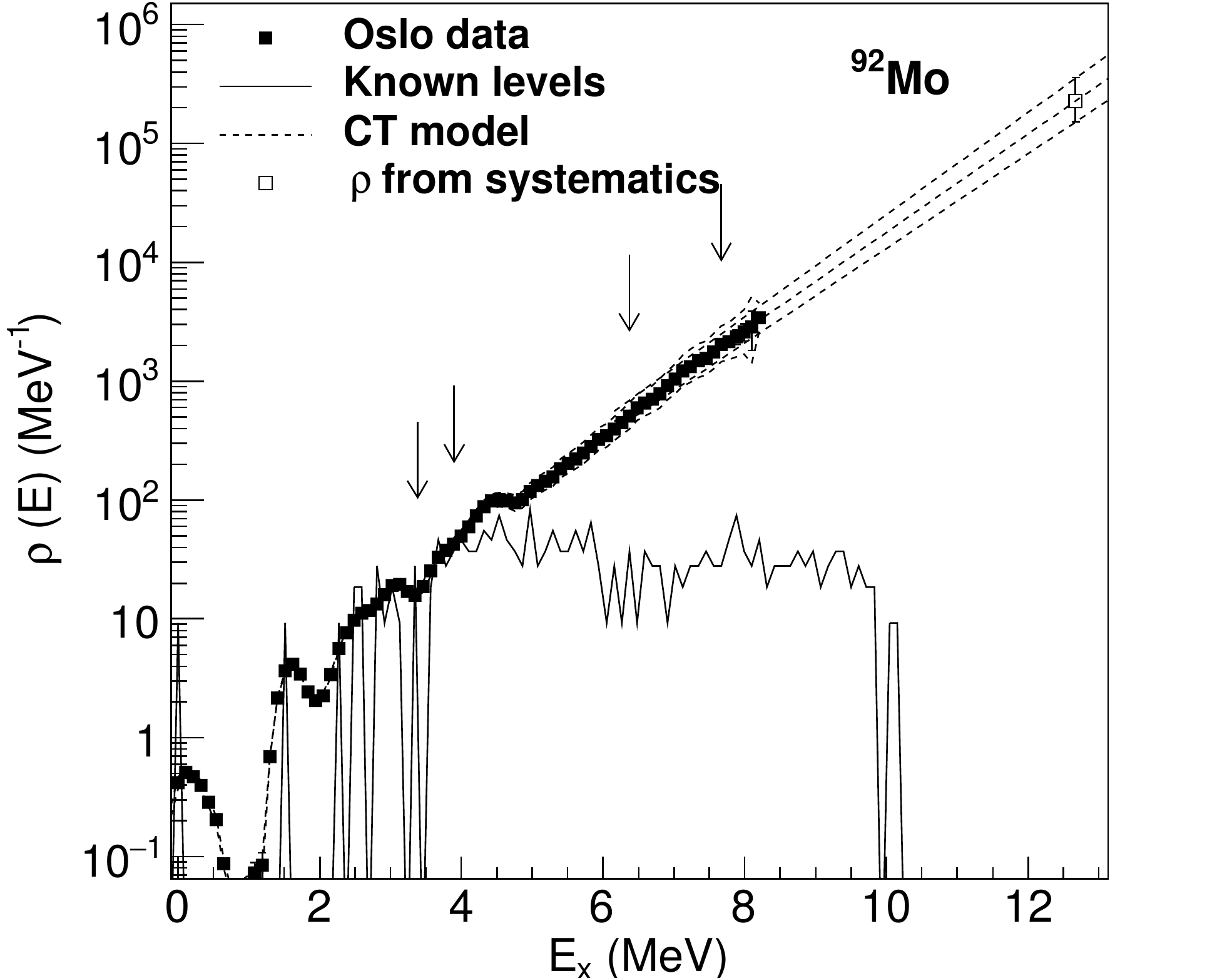}
\caption{The experimentally extracted upper and lower limits of $\rho(E)$ for $^{92}$Mo. The bin width of $E_x \approx 0.1$ MeV \label{fig:fig2}.}
\end{figure}

To estimate the systematic uncertainty in the normalization procedure, a set of normalizations were used. The upper normalization value at $\rho(S_n)$ was obtained by increasing the Back Shifted Fermi Gas global systematics with the parametrization of Ref.\cite{PhysRevC.72.044311,PhysRevC.73.049901} by 16$\%$ to fit the experimental values at $S_n$ for the Mo isotopes \cite{PhysRevC.88.015805}. The middle normalization was chosen to be compatible with the value obtained using the spin cutoff parameter calculated according to Ref.\cite{PhysRevC.80.054310} and increased by 80$\%$ so that the model agrees with the experimental value for the best studied Mo-isotope $^{96}$Mo \cite{PhysRevC.88.041305}. The lowest normalization was found by using the same spin cutoff model as for the middle normalization and selecting the lowest value of $\rho(S_n)$ that gives a normalization of the $\gamma$SF consistent with data taken for $E_x > S_n$. As for other Mo isotopes \cite{mo2, mo3}, the NLD above $\approx$ 5 MeV is well described by the Constant-Temperature formula, $\rho_{CT}(E_x)=\frac{1}{T}e^{(E_x-E_0)/T}$, where $T$ is the temperature and $E_0$ is the energy shift \cite{consttemp,ctmgc}. Therefore, the $\rho_{CT}$ model is used for extrapolating up to $\rho(S_n)$. The three normalizations of the NLD, $\rho(S_n)=2.28^{+1.27}_{-0.76} \cdot 10^5$ MeV$^{-1}$, of $^{92}$Mo are shown in Fig. \ref{fig:fig2}. The results will be published online \cite{oclcomp}. 
 The $\gamma$SF is deduced from $\mathfrak{T}(E)$ by $f(E_{\gamma}) = \mathfrak{T}(E) / 2\pi{E_{\gamma}}^3$, where $f(E_{\gamma})$ is the $\gamma$SF. For the normalization of the $\gamma$SF, systematics for the Mo isotopes given in Ref.\cite{PhysRevC.88.041305,PhysRevC.88.015805} were applied, as well as the requirement that the $\gamma$SF below $S_n$ should be compatible with data from other experiments above $S_n$. It has long been suspected that the neutron strength only accounts for part of the total giant dipole resonance (GDR) strength \cite{Beil1974427,SHODA1975397}. According to the Thomas-Reiche-Kuhn (TRK) sum rule \cite{trk1,trk2,trk3} for the GDR strength, $\int \sigma_{\gamma}(E)dE = 60 NZ/A$  MeV mb,  the total strength of the GDR varies little within a given isotopic chain. Therefore, also $(\gamma,n)$ data for neighbouring Mo isotopes were used as guide.

\begin{figure}[tb]
\includegraphics[width=0.5\textwidth]{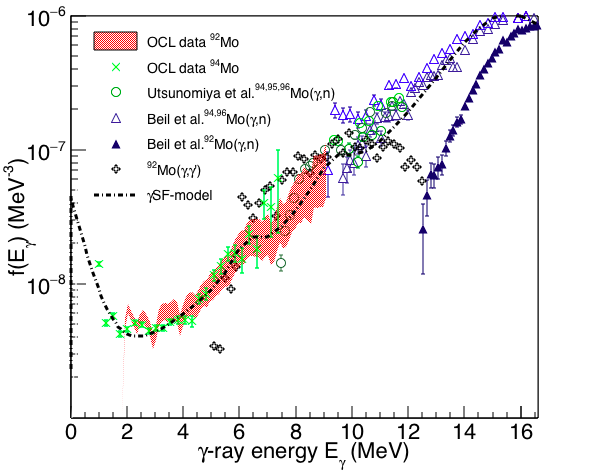}
\caption{The upper and lower limit for $f(E_{\gamma})$ compared to $(\gamma,\gamma')$-data from ELBE \cite{Rusev2009}, $(\gamma,n)$-data for $^{95,96}$Mo \cite{PhysRevC.88.015805} and $(\gamma,n)$-data for $^{92,94,96}$Mo \cite{Beil1974427,mo4}. The renormalized $^{94}$Mo OCL data is also shown \cite{mo3,mo4}. The $\gamma$SF-model used as input to TALYS is also shown.\label{fig:fig3}}
\end{figure}

Combining upper and lower limits on the $\langle\Gamma_{\gamma}(S_n)\rangle$ values found by studying the systematics of the Mo-isotopes with the upper and lower normalizations of the NLD respectively provides a set of  normalizations for the $\gamma$SF, as shown in Fig. \ref{fig:fig3}. The three sets of normalizations for the NLD and $\gamma$SF are given in Tab.\ref{tab:normpar}. 
\begin{table}[tb]
\centering
\caption{Normalization parameters for $\rho(E_x)$ and $f(E_{\gamma})$. \label{tab:normpar}}
\begin{ruledtabular}
\begin{tabular}{lccc}
 Parameter & middle & upper &  lower \\
\hline
 $\rho(S_n)$ (10$^5$ MeV$^{-1}$)&  2.28 &  3.55 &  1.52   \\
 $D_0$ (eV)&  33& 27  & 48   \\
$\langle\Gamma_{\gamma}(S_n)\rangle$ (meV)&270  & 290 &  250 \\
 $\sigma$ & 4.4  & 5.7 & 4.2  \\
\end{tabular}
\end{ruledtabular}
\end{table}

For the other Mo isotopes where the $\gamma$SF has been studied, a low-energy enhancement of the  $E_{\gamma} <$ 3 MeV has been observed for $^{93-98}$Mo \cite{mo3,Wiedeking}. For the present data set on $^{92}$Mo, the same feature is present. The low-energy upbend has been shown to be of dipole nature \cite{feprl}; however, the electromagnetic character has not been experimentally determined. At present, there exist two theoretical predictions: in the work of Ref.\cite{elena}, presenting calculations on the $^{94,96,98}$Mo $\gamma$SF within the framework of the quasi-particle random-phase approximation, it is claimed that the upbend is of electric character. On the other hand, shell-model calculations \cite{sch13M1} indicate a strong low-energy increase in the M1 component of the $\gamma$SF for $^{94,95,96}$Mo. In this work, the low-energy behaviour of the $^{92}$Mo $\gamma$SF has been studied by means of the shell-model code RITSSCHIL \cite{zwa85}. The calculations were carried out using a model space composed of the $\pi(0f_{5/2}, 1p_{3/2}, 1p_{1/2}, 0g_{9/2})$ proton and $\nu(0g_{9/2}, 1d_{5/2})$ neutron orbits relative to a $^{66}$Ni core. This configuration space was also applied in our earlier study of $M1$ and $E2$ strength functions in $^{94,95,96}$Mo and $^{90}$Zr \cite{sch13M1,sch14E2}. 

The calculations included the lowest 40 states each for spins from $J$ = 0 to 10. Reduced transition strengths $B(M1)$ were calculated for all possible transitions with spins $J_f = J_i, J_i \pm 1$. This resulted in more than 23700 $M1$ transitions for each parity, which were sorted into 100 keV bins according to their transition energy. 

The $M1$ $\gamma$SFs were deduced by using the relation $f_{M1}(E_\gamma) = 16\pi/9$ $(\hbar c)^{-3}$ $\overline{B}(M1,E_\gamma)$ $\rho(E_i)$. They were calculated by multiplying the ${B(M1)}$ value in $\mu^2_N$ of each transition with $11.5473 \times 10^{-9}$ times the level density at the energy of the initial state $\rho(E_i)$ in MeV$^{-1}$ and deducing averages in transition energy. The level densities $\rho(E_i,\pi)$ were determined by counting the calculated levels within energy intervals of 1 MeV for the two parities separately. The $\gamma$SF obtained for the two parities were subsequently added. When calculating the $\gamma$SF, gates were set on the excitation energy, 7 MeV $\leq E_x \leq$ 10.2 MeV, corresponding to those applied in the analysis of the experimental data. 

\begin{figure}[tb]
\includegraphics[width=0.5\textwidth]{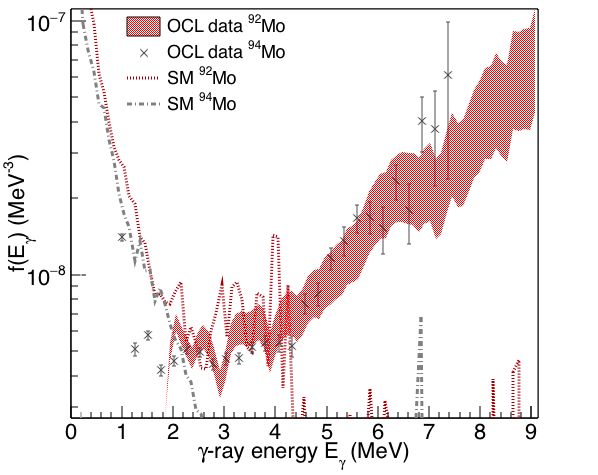}
\caption{Shell model calculations for $^{92,94}$Mo shown together with the $^{92,94}$Mo OCL data \cite{mo4}.\label{fig:94mo}}
\end{figure}

The value of $B(E2) = 146 e^2$fm$^4$ calculated for the
$2^+_1 \rightarrow 0^+_1$ transition in $^{92}$Mo using
effective charges of $e_\pi = 1.5 e$ and $e_\nu = 0.5 e$
has to be compared with an experimental value of
$B(E2) = 206(12) e^2$fm$^4$ \cite{bag12}. The calculated
value is closer to the experimental one than the corresponding
value in the neighboring heavier isotope $^{94}$Mo \cite{sch14E2},
thus reflecting the little collectivity of the $N$ = 50 nuclide.
As seen in Fig.~\ref{fig:94mo}, the calculated strength in $^{92}$Mo
exceeds the experimental strength in the neighbor $^{94}$Mo. It is also
somewhat higher than the calculated one for $^{94}$Mo at low energy, as can also be seen from Fig. \ref{fig:94mo}.
The calculations nevertheless provide a viable explanation of the upbend.

Although the upbend predicted by the shell-model calculations is considerable at low $\gamma$-ray energies, it is not expected to contribute much to the average total radiative width $\left< \Gamma_\gamma \right>$ at the neutron or proton separation energy, because it is situated at very low $E_\gamma$ energies compared to the separation energies of $^{92}$Mo ($S_n = 12.67$ MeV, $S_p = 7.46$ MeV). For nuclei with small $S_n,S_p$ values the upbend would be expected to contribute significantly to $\left< \Gamma_\gamma \right>$, and hence influence the astrophysical reaction rates (see e.g. Ref.~\cite{acl2010}). However, the full $\gamma$SF up to the particle thresholds is undisputably of great importance for the reaction rates.

\begin{figure}[tb]
\includegraphics[width=0.5\textwidth]{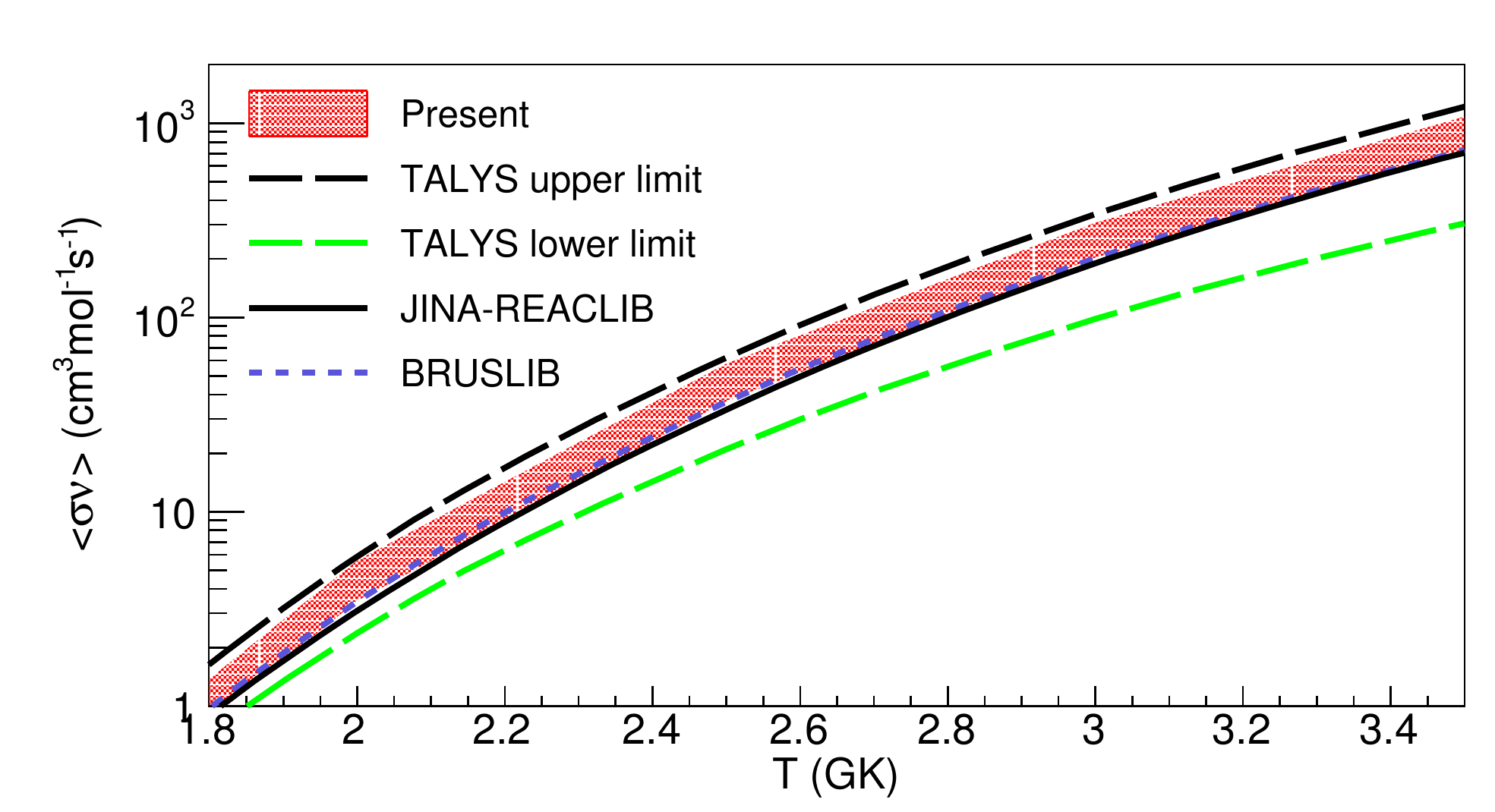} 
\includegraphics[width=0.5\textwidth]{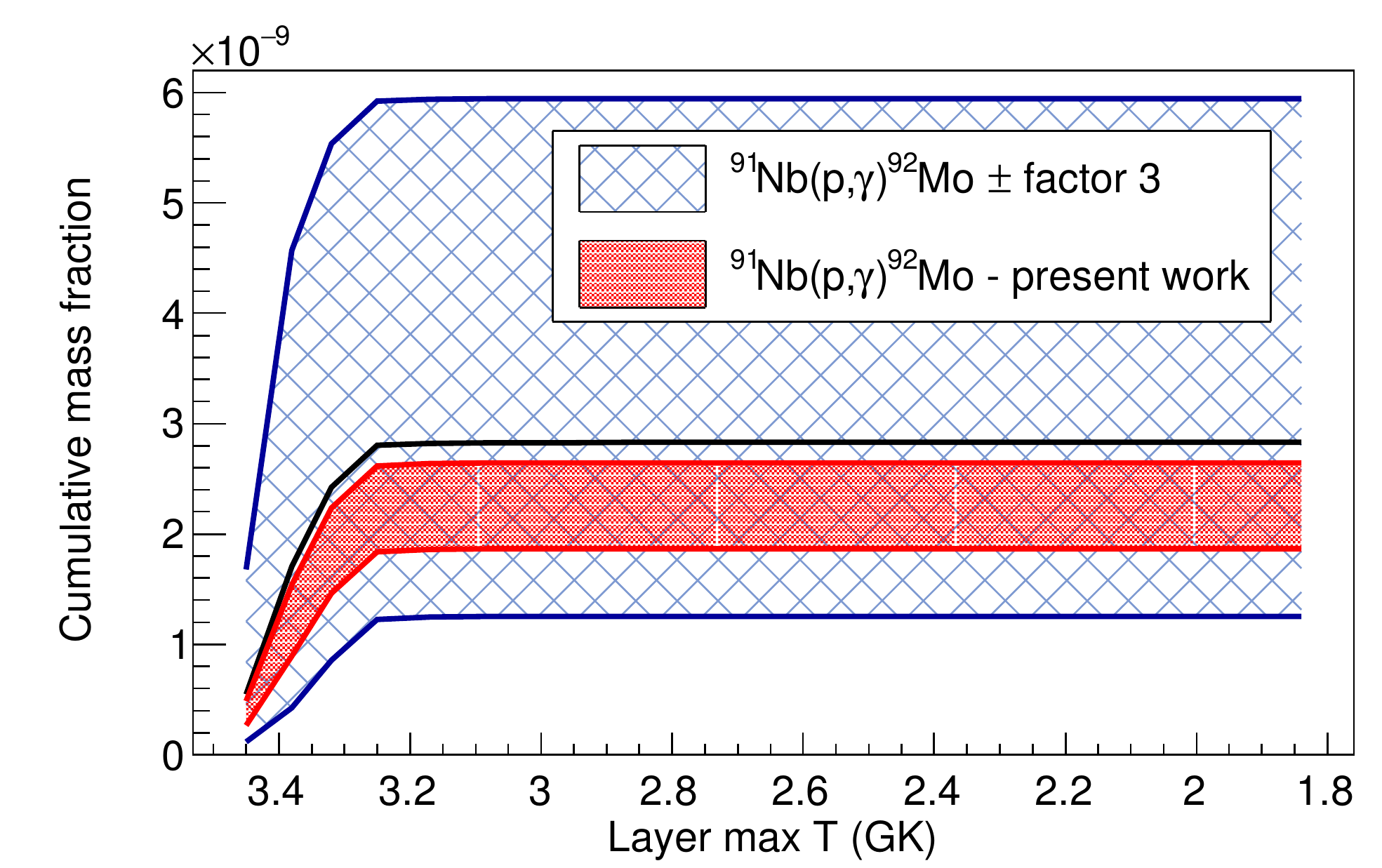}
\caption{Upper panel: Comparison between the data-guided TALYS 1.6 calculations and theoretical predictions for the astrophysical reaction rates. Lower panel: The accumulation of $^{92}$Mo in different zones for the reaction rate from the JINA REACLIB $\pm$ factor 3 compared to the same calculations using the present experimental results as input. \label{fig:fig4}}
\end{figure}

The astrophysical reaction rates for the $^{91}$Nb(p,$\gamma$)$^{92}$Mo reaction were calculated with TALYS 1.6  \cite{talys, koning12}, using input guided by our experimental results for $^{92}$Mo. That the nuclei can exist in various excited states in a stellar environment, and in particular the 104.6 keV isomeric state of $^{91}$Nb is taken into account in the astrophysical calculations of TALYS. The default global optical model parameters were used for the lower limits \cite{KD03} and the semi-microscopic nucleon-nucleus spherical optical model (JLM) for the upper limits \cite{JLMOMP,Capote20093107}. The TALYS input for the NLD and $\gamma$SF for $^{92}$Mo were adjusted to match closely the experimental NLD and $\gamma$SF for $^{92}$Mo. The generalized Lorentzian model of Kopecky and Uhl \cite{PhysRevC.41.1941} with RIPL-3 parameters for the GDR strength as the starting point and a constant temperature adjusted to fit with $(\gamma$,n) other experimental data above $S_n$ and the $\gamma$SF below $S_n$ was used. In addition, two standard Lorentzian resonances (Res 1 and Res 2) were included to replicate the experimental results.  Finally, an exponential function $f(E_{\gamma})^{upbend} = C \exp(\eta E_{\gamma})$ was adjusted to fit the low energy upbend of the OCL data for $^{94}$Mo \cite{mo3,mo4}. The inclusion of the upbend accounts for $0-3\%$ of the total rate for the temperatures investigated in this work (0-10 GK). The total $\gamma$SF of $^{92}$Mo used as input to the TALYS calculations is given by Eq. \ref{eq:eq1} with the parameters provided in Tab.\ref{tab:respar}. 
\begin{equation} \label{eq:eq1}
f(E_{\gamma}) = f^{GDR} + f^{Res 1}  + f^{Res 2} +  f^{upbend}
\end{equation}  
The resulting $\gamma$SF input for TALYS is shown in Fig. \ref{fig:fig3}. The experimentally constrained reaction rate of the $^{91}$Nb$(p,\gamma)^{92}$Mo reaction is shown in Fig. \ref{fig:fig4} (upper panel). The temperature range for this reaction in typical astrophysical sites for the p-process is 1.8 - 3.4 GK \cite{Rauscher2013}. The results of the present work are compared to TALYS calculations using standard NLD and $\gamma$SF input. The TALYS upper limit corresponds to the Generalised superfluid NLD model \cite{NLD3_1,NLD3_2} and the Brink-Axel $\gamma$SF \cite{AxelBrink1,AxelBrink2}, while the TALYS lower limit is obtained with microscopic level densities \cite{NLD6} and Hartree-Fock BCS tables for the $\gamma$SF \cite{Capote20093107}. The present experimental lower-limit result is in good agreement with the theoretical calculations. Fig. \ref{fig:fig4} (upper panel) includes the reaction rate from the two commonly used reaction libraries JINA REACLIB\cite{jina} and BRUSLIB\cite{bruslib}. 
The present experimental result provides a strong experimental constraint on the reaction rate.
\begin{table}[tb]
\centering
\caption{Resonance and NLD parameters used as input to TALYS 1.6. \label{tab:respar}}
\begin{ruledtabular}
\begin{tabular}{lcccc}
 Resonance & Parameter& middle & upper & lower   \\
\hline
GDR& $E$ [MeV]& 16.04 &  16.04& 16.03   \\
 & $\sigma$ [mb]& 188 &  188& 188  \\
& $\Gamma$ [mb]&  4.5& 4.6 & 4.2  \\
& $T$ [MeV] &  0.64& 0.59 & 0.59  \\
\hline
Res 1 & $E$ [MeV]&  9.4& 9.5 & 9.4   \\
 &$\sigma$ [mb]&  4.7&  9.2& 3.2  \\
&$\Gamma$ [mb]&1.5  & 1.7 & 1.4  \\
\hline
Res 2 &  $E$ [MeV]& 6.3 & 6.4  & 6.3   \\
 &$\sigma$ [mb]& 0.72  & 0.79 & 0.42  \\
&$\Gamma$ [mb]& 0.57 & 0.76 & 0.67  \\
\hline
Upbend& $C$ [MeV$^{-1}$]& 4.3$\cdot10^{-8}$ &4.3$\cdot10^{-8}$ & 4.3$\cdot10^{-8}$  \\
& $\eta$ [MeV$^{-3}$]&  -1.9 & -1.9 & -1.9   \\
\hline
CT NLD&$T$ [MeV] &  1.10 & 1.16 & 1.06  \\
&$E_0$ [MeV]&0.79 & 0.64 & 0.9 \\
\end{tabular}
\end{ruledtabular}
\end{table}

The reaction rate extracted in this work was used in reaction network calculations for the scenario of a p-process taking place in a type II supernova explosion as the shock front passes through the O-Ne layer of a 25$ M _ {\odot} $ star. The astrophysical calculations were performed using the post processing code available in NUC NET tools \cite{nucnet}, a suite of nuclear reaction codes developed at Clemson University. The calculations were performed in a multilayer model (14 layers) using the seed distribution of a pre-explosion 25$ M _ {\odot} $ star. The seed distribution and temperature and density profiles were taken from Ref.\cite{psensitivity}. For reaction rates other than the one studied here the JINA REACLIB input was used.

In these calculations, the $^{92}$Mo mass fraction was extracted for each layer, which is a measure of the calculated abundance for this isotope. The cumulative mass fraction of $^{92}$Mo is shown in Fig. \ref{fig:fig4} (lower panel). The graph starts with the inner layer (highest temperature) and the total mass fraction of $^{92}$Mo accumulates moving outward to layers with lower maximum temperature. Only the layers with the highest maximum temperature contribute to the accumulation of $^{92}$Mo. The black line corresponds to the cumulative mass fraction of $^{92}$Mo using standard reaction rates from the JINA REACLIB. Varying the rate of the $^{91}$Nb(p,$\gamma$)$^{92}$Mo reaction by a factor of 3 up and down (same factor used as standard in Ref. \cite{psensitivity}) changes the mass fraction as indicated by the checkered area. Using the experimental upper and lower limits from the present work, the mass fraction uncertainty is significantly reduced, as shown by the hatched area. The present result provides a stringent constraint on the last unmeasured reaction related to the nucleosynthesis of $^{92}$Mo, and reinforces the conclusion that the underproduction of $^{92}$Mo cannot be attributed to the nuclear physics input.  Indeed, since the entire uncertainty band for the cumulative production of $^{92}$Mo lies below the JINA Reaclib rate, this new analysis may somewhat exacerbate the $^{92}$Mo underproduction problem in astrophysical models of the p-process.

In summary, the experimentally extracted NLD and $\gamma$SF of $^{92}$Mo have been used as input to TALYS calculations for the $^{91}$Nb$(p,\gamma)^{92}$Mo reaction. This work provides the first stringent experimental constraint for this remaining part of the nuclear reaction puzzle of the nucleosynthesis of $^{92}$Mo. We conclude that the reason for the underproduction of $^{92}$Mo is not related to the $^{91}$Nb(p,$\gamma$)$^{92}$Mo cross section input to astropysical models of the p-process.

\begin{acknowledgments}
We would like to give special thanks to J.C. M\"uller, A. Semchenkov, and J.C. Wikne for providing the high quality beam and excellent experimental conditions. Lawrence Berkeley National Laboratory is thanked for lending us the $^{92}$Mo target. G.M.T.  gratefully acknowledges funding of this research from the Research Council of Norway, Project Grant No. 222287. A.C.L acknowledges funding from ERC-STG-2014 grant agreement no. 637686. This work was supported by the National Science Foundation under Grants No PHY1102511 (NSCL), No. PHY 1430152 (JINA-CEE) and No PHY 1350234 (CAREER). This work was performed under the auspices of the US Department of Energy DE-AC52-07NA27344 (LLNL) and DE-AC02-05CH11231 (LBNL).
\end{acknowledgments}

\bibliography{mybibfile}

\end{document}